\documentclass[aps,apl,twocolumn,superscriptaddress]{revtex4}
\usepackage{epsfig}
\usepackage[usenames]{color}
\DeclareMathAlphabet{\mathitb}{OT1}{cmr}{bx}{sl}

\begin{document}

\renewcommand{\thefootnote}{\fnsymbol{footnote}}

\title{Electron Spin Excited States Spectroscopy in a Quantum Dot Probed by QPC Back-action }

\author{HaiOu Li}
\author{Ming Xiao}
\email{maaxiao@ustc.edu.cn}
\author{Gang Cao}
\author{Cheng Zhou}
\author{RuNan Shang}
\author{Tao Tu}
\author{GuangCan Guo}
\affiliation{Key Laboratory of Quantum Information, Chinese Academy of Sciences, University of Science and Technology of China, Hefei 230026, People's Republic of China}
\author{HongWen Jiang}
\affiliation{Department of Physics and Astronomy, University of California at Los Angeles, 405 Hilgard Avenue, Los Angeles, CA 90095, USA}
\author{GuoPing Guo}
\email{gpguo@ustc.edu.cn}
\affiliation{Key Laboratory of Quantum Information, Chinese Academy of Sciences, University of Science and Technology of China, Hefei 230026, People's Republic of China}
\date{\today}

\begin{abstract}
The quantum point contact (QPC) back-action has been found to cause non-thermal-equilibrium excitations to the electron spin states in a quantum dot (QD). Here we use back-action as an excitation source to probe the spin excited states spectroscopy for both the odd and even electron numbers under a varying parallel magnetic field. For a single electron, we observed the Zeeman splitting. For two electrons, we observed the splitting of the spin triplet states $|T^{+}\rangle$ and $|T^{0}\rangle$ and found that back-action drives the singlet state $|S\rangle$ overwhelmingly to $|T^{+}\rangle$ other than $|T^{0}\rangle$. All these information were revealed through the real-time charge counting statistics.
\end{abstract}

\maketitle

The spin states of few electrons in quantum dots (QD) have been demonstrated as potential candidates for qubits \cite{Spin-Qubits-Kouwenhoven, Spin-Qubits-Petta}. The Zeeman states for odd number of electrons and the spin singlet-triplet states for even number of electrons form the basis of qubit operation and detection. Also we know that any detection to the qubit states would necessarily have back-action \cite{Backaction-Kouwenhoven, Backaction-Dephase}. Recently we found that the back-action from a quantum point contact (QPC) would excite the spin singlet to triplet states and degrade the fidelity of qubit operation \cite{Ming-RTS-ST}. However, this study was limited to the non-degenerate spin states at zero magnetic field. 

To know the influence of back-action on all relevant spin states would be very helpful for spin based qubits. In this work we applied a parallel magnetic field to lift the spin states degeneracy and explored their spectroscopy. By studying the non-equilibrium part of the real-time charge counting statistics under strong back-action condition, we observed the excitation from an up to a down spin state for one single electron, and the excitation from the spin singlet state $|S\rangle$ to triplet states $|T^{+}\rangle$ and $|T^{0}\rangle$ for two electrons. We revealed the linear dependence of the exchange energy on the magnetic field. More importantly, we found that back-action overwhelmingly excites $|S\rangle$ to $|T^{+}\rangle$ state. This implies that spin qubits based on $|S\rangle - |T^{0}\rangle$ are more immune of back-action than $|S\rangle - |T^{+}\rangle$.

\begin{figure}[t]
\begin{center}
\epsfig{file=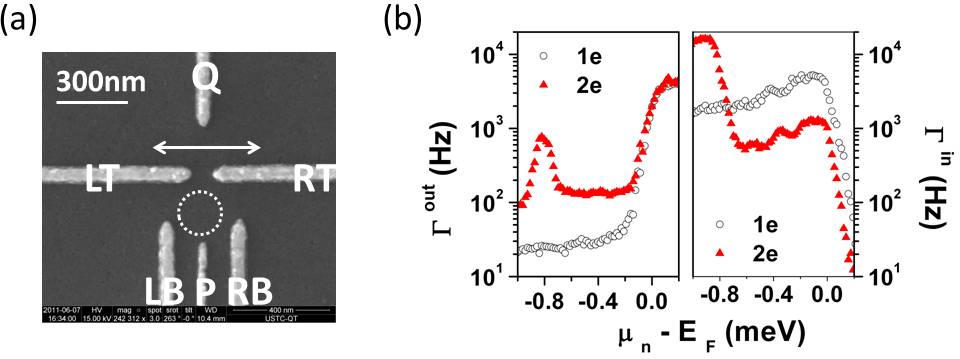, width=1\linewidth, angle=0}
\end{center}
\vspace{-7.5mm}
\caption{(a) A SEM picture showing the geometry of our sample. The dotted circle is the QD and the arrowed line indicates the QPC channel.  (b) RTS statistics $\Gamma^{out}$ and $\Gamma^{in}$. Black open circles for $1e$ and red closed triangles for $2e$.}
\label{Figure1}
\end{figure}

The GaAs sample is the same as in earlier experiments \cite{Ming-RTS-ST, Ming-RTS-Backaction}. Fig. \ref{Figure1} is a scanning electron microscopy (SEM) image of the QD-QPC structure. The QD is configured to be in thermal equilibrium with the electron reservoir on the right side. Random telegraph signal (RTS), the single electron tunneling back and forth between the QD and the right side reservoir due to thermal fluctuations, has been recorded and analyzed. Fig. \ref{Figure1} (b) shows the RTS statistics $\Gamma^{out}$ and $\Gamma^{in}$ for the last two electrons. As we found earlier, the QPC back-action caused severe non-thermal-equilibrium effect. For the $(n-1)e \leftrightarrow ne$ tunneling, the back-action drives the $n^{th}$ electron out of the QD even when the electron addition energy $\mu_{n} <<$ the reservoir Fermi level $E_{F}$ \cite{Ming-RTS-Backaction}. As a result, $\Gamma^{out}$ shows a strong saturation tail for both $1e$ and $2e$ when $\mu_{n} - E_{F}<<0$. What's more, those slowly relaxing excited states provide additional tunneling channels and give rise to extra features \cite{Ming-RTS-ST}. For $2e$, we see a strong side peak in $\Gamma^{out}$ and an extra elevated plateau in $\Gamma^{in}$ at about  $-0.8meV$, arising from the spin singlet-triplet excitation. For $1e$, no extra feature is seen since the spin up and down states are degenerate and indistinguishable.

\begin{figure}[t]
\begin{center}
\epsfig{file=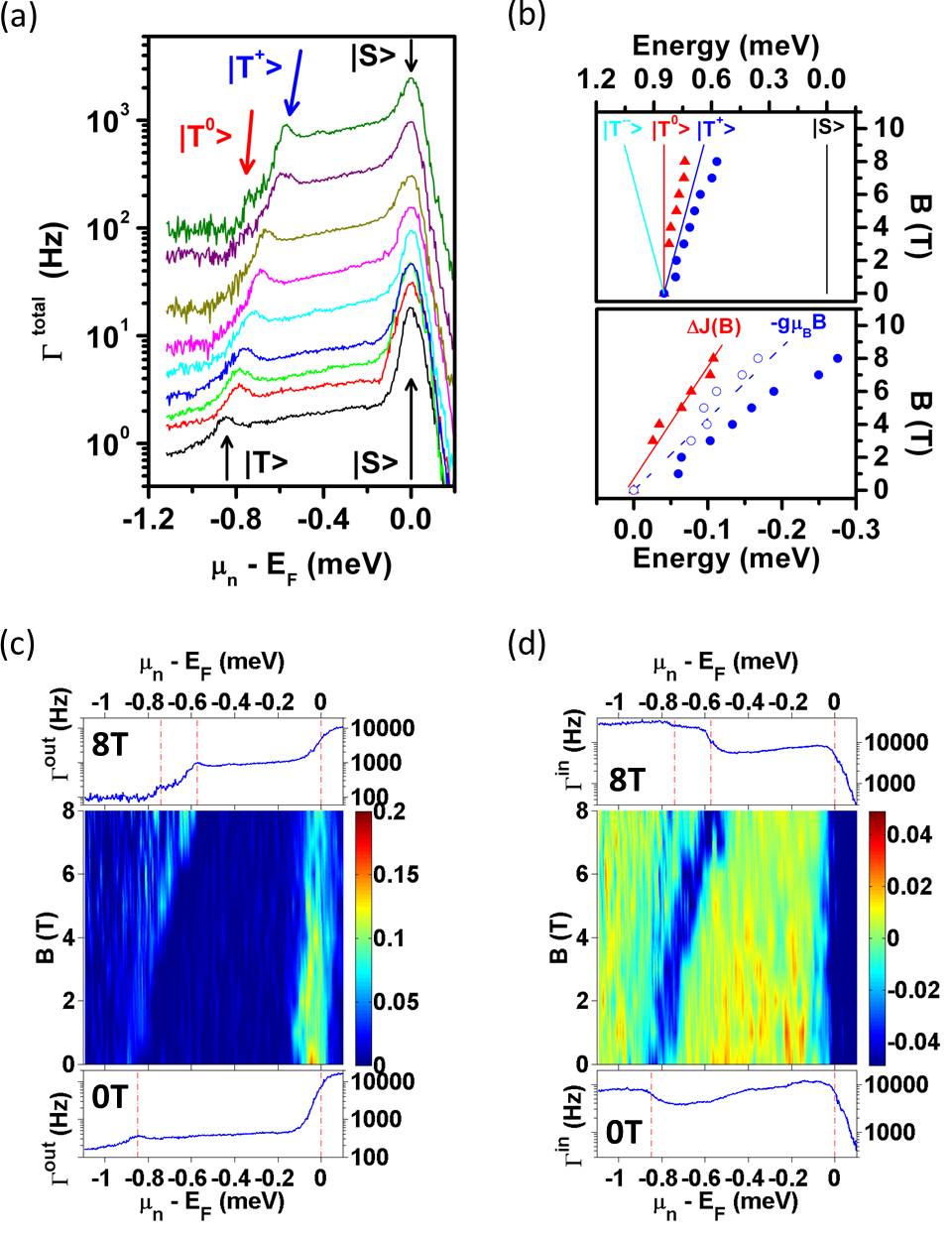, width=1\linewidth, angle=0}
\end{center}
\vspace{-7.5mm}
\caption{(a) $\Gamma^{total}$ for $2e$ at different magnetic field. From bottom to up, $B$ increases from $0T$ to $8T$ by $1T$ each step. All traces except $8T$ are shifted on the y-axis for comparison. (b) Up: extracted energy of $|T^{0}\rangle$ and $|T^{+}\rangle$ with respect to $|S\rangle$. Red triangles for $|T^{0}\rangle$ and  blue circles for $|T^{+}\rangle$. Solid lines are the predicted energy for all the spin singlet and triplet states. Down: Red triangles are the extracted energy shift of $|T^{0}\rangle$ with magnetic field, which represents the shift of the exchange energy, i.e, $\Delta J(B)$. Red line is the linear fitting. Blue closed circles are the energy shift of $|T^{+}\rangle$. Blue open circles are the energy difference between $|T^{+}\rangle$ and $|T^{0}\rangle$, which agrees well with the predicted Zeeman splitting assuming g-factor is $0.4$ in GaAs, as represented by the blue dashed line. (c) 3D plot of $\Gamma^{out}$ versus magnetic field and electron energy. On the top and bottom, an original trace at $B=8T$ and $0T$ is respectively presented. (d) 3D plot of $\Gamma^{in}$.}
\label{Figure2}
\end{figure}

Now we apply a parallel magnetic field to lift the energy degeneracy between the two Zeeman states for the odd number of electrons, and between the three triplet states for the even electron numbers. This will enable us to probe the detailed spin excited states spectroscopy and study the effect of back-action on each individual spin states. Fig. \ref{Figure2} (a) shows the total tunneling rate $\Gamma^{total}$ for $2e$ with varying magnetic field strength $B$. The black line at the bottom is at $B=0T$. Due to the strong back-action, we see both the ground state  $|S\rangle$ and the excited states $|T\rangle$, located at $\mu_{2} - E_{F}=0$ and $-0.85meV$, respectively. At this point, we don't know $|T\rangle$ refers to which one of the three energy degenerate configurations $|T^{+}\rangle$, $|T^{0}\rangle$ or $|T^{-}\rangle$. As we increase the magnetic field, the peak at $-0.85meV$ shifts closer to the singlet state $|S\rangle$, at a speed comparable to that of the Zeeman splitting. This suggests that the $|T^{+}\rangle$ state plays the major role. The $|T^{0}\rangle$ state, which stays more constant, is also visible at large magnetic field ($B\geq3T$). But its peak amplitude is much smaller than that of $|T^{+}\rangle$. The third state $|T^{-}\rangle$ is not visible in this experiment. This phenomena is also presented in Fig. \ref{Figure2} (c) and (d), which are the 3D diagrams for the numerical derivative of $\Gamma^{out}$ and $\Gamma^{in}$.  The triplet state splits into two with increasing magnetic field. The one obviously shifting towards the singlet state is $|T^{+}\rangle$, and the other one staying more constant is $|T^{0}\rangle$.

In Fig. \ref{Figure2} (b) we explicitly extracted the energy of each states with respect to the singlet energy, as functions of the magnetic field. On the upper part of the figure the blue circles are the energy of $|T^{+}\rangle$ and the red triangles are that of $|T^{0}\rangle$. Only at $B \geq 3T$ we are able to collect data for $|T^{0}\rangle$. The solid lines show the standard picture of the spin singlet-triplet energy spectrum. At $B=0$, the three triplet states are degenerate with a certain separation from $|S\rangle$, called exchange energy $J$. With increasing $B$, $|T^{0}\rangle$ is assumed invariant while  $|T^{+}\rangle$ and  $|T^{-}\rangle$ shift by the amount of  $\pm g\mu_{B}B$ respectively. Experimentally, we found that the separation between $|T^{+}\rangle$ and  $|T^{0}\rangle$ fits very well with the predicted Zeeman splitting using a g-factor $-0.4$ in GaAs, as indicted by the blue open circles and blue dashed line in the lower part of Fig. \ref{Figure2} (b). This quantitatively proves that the two states we have seen are the two different spin excited states. 

What's more, we noticed that the energy of  $|T^{0}\rangle$ is not exactly invariant under magnetic field. It slightly decreases with field, which means that the exchange energy $J(B)$ is actually a function of magnetic field. In the lower part of Fig. \ref{Figure2} (b) we presented $\Delta J(B)$, defined as $J(B)-J(0)$. The linear fitting gives a slope $(-0.015\pm0.001) meV/T$. If we don't consider the decrease of exchange energy with field, then the extracted energy shift of $|T^{+}\rangle$ itself is too large to be explained as the Zeeman splitting, shown as the blue closed circles in the lower part of Fig. \ref{Figure2} (b). Thus the field dependence of $J(B)$ must be taken into account. The decrease of exchange energy could be due to the confinement of the electron wavefunctions by magnetic field, which isolates the two electrons and decreases their interaction.

\begin{figure}[t]
\begin{center}
\epsfig{file=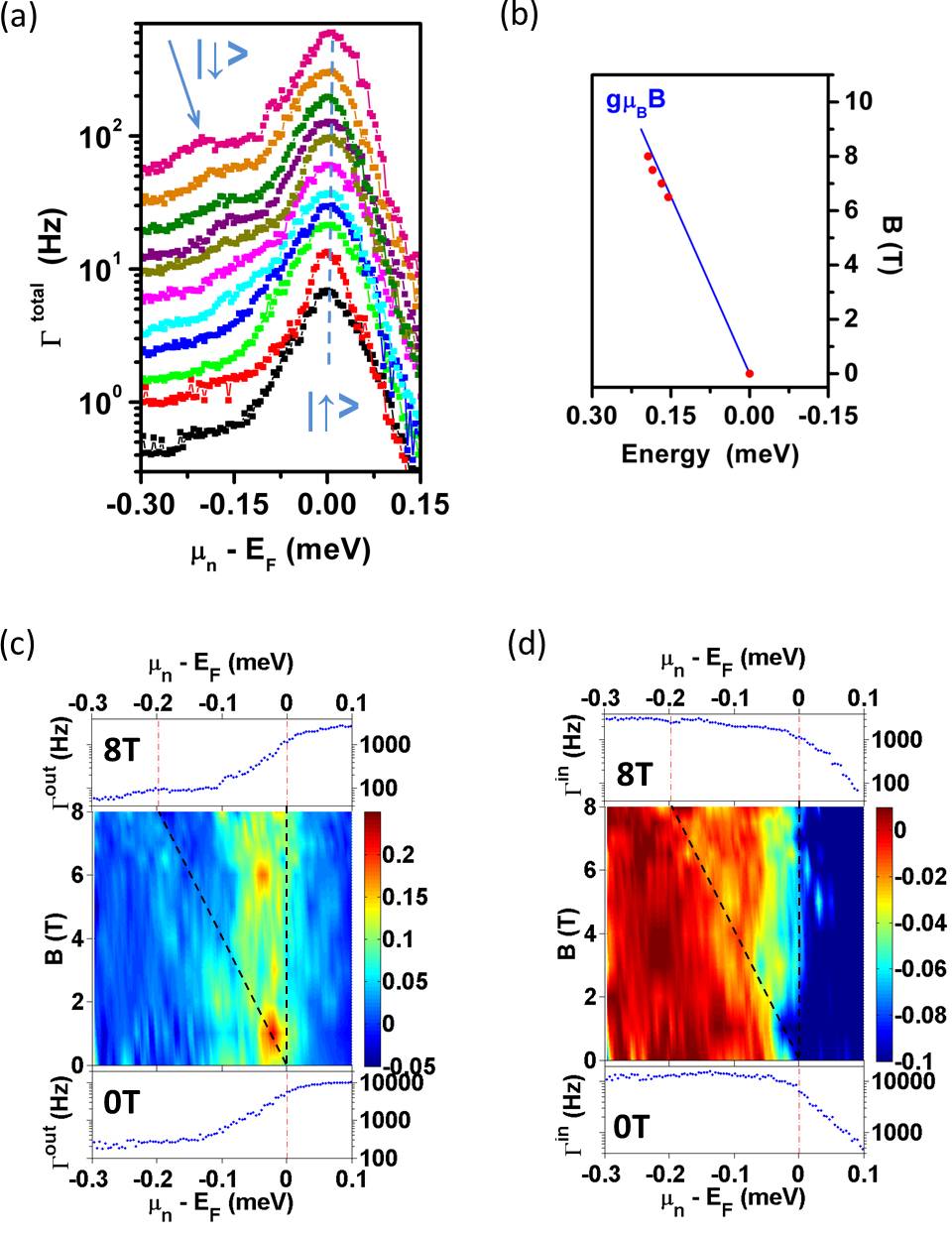, width=1\linewidth, angle=0}
\end{center}
\vspace{-7.5mm}
\caption{(a) $\Gamma^{total}$ for $1e$ at different magnetic field. From bottom to up, $B$ varies from $0T$ to $8T$. $B$ increases from $0T$ to $6T$ by $1T$ each step, and then increases from $6.5T$ to $8T$ by $0.5T$ each step. All traces except $8T$ are shifted on the y-axis for comparison. (b) Extracted energy of $\mid\downarrow\rangle$ with respect to $\mid\uparrow\rangle$. The solid line is the predicted Zeeman splitting. (c) 3D plot of $\Gamma^{out}$ versus magnetic field and electron energy. On the top and bottom, an original trace at $B=8T$ and $0T$ is respectively presented. (d) 3D plot of $\Gamma^{in}$.}
\label{Figure3}
\end{figure}

One striking finding in the above experiment is that the $|T^{+}\rangle$ state is much more prominent than $|T^{0}\rangle$. No matter for $\Gamma^{total}$, $\Gamma^{out}$ or $\Gamma^{in}$, the $|T^{+}\rangle$ state is always the major structure while $|T^{0}\rangle$ is barely visible until it is separated from $|T^{+}\rangle$ by a large enough magnetic field. Using the phenomenological model that we developed to analyze the back-action driven tunneling rates \cite{Ming-RTS-ST}, we could quantitatively compare the excitation rate from $|S\rangle$ to $|T^{+}\rangle$ and $|T^{0}\rangle$. Simulation on the data taken at $8T$ first reveals that the back-action driven rate for the electrons to tunnel out of the QD is $0.4\%$ of the maximum RTS tunneling rate for the ground state $|S\rangle$ ($\Lambda_{out}=30Hz$ and $\Gamma_{S}^{*}=8kHz$). $\Gamma^{*}$ for the two singlet states $|T^{+}\rangle$ and $|T^{+}\rangle$ varies a little ($\Gamma_{T^{+}}^{*}=15kHz$ and $\Gamma_{T^{0}}^{*}=12kHz$), and their relaxation time also shows small change ($1.2ms$ and $1.5ms$). However, the back-action driven excitation rate from $|S\rangle$ to $|T^{+}\rangle$ is $6$ times large as that from $|S\rangle$ to $|T^{0}\rangle$: $\Lambda_{ST^{+}}=600Hz$ and $\Lambda_{ST^{0}}=100Hz$. It is this big difference that makes the $|T^{+}\rangle$ state dominate the excited states spectroscopy.

Thus, we made the conclusion that QPC back-action prefers driving the spin singlet $|S\rangle$ up to $|T^{+}\rangle$ other than $|T^{0}\rangle$ state. It is uaually assumed that due to the spin-phonon selection rules, the $|T^{0}\rangle-|S\rangle$ interaction is suppressed in the lowest order, which makes the relaxation from $|T^{0}\rangle$ to $|S\rangle$ longer than the other two triplet states \cite{Spin-Relaxation-Selection -Rules}. On the other hand, this also means that the phonon mediated back-action between $|T^{0}\rangle$ and $|S\rangle$ is weaker. This could possibly explain our observation. Either $|T^{0}\rangle-|S\rangle$ or $|T^{+}\rangle-|S\rangle$ has been chosen as the basis for qubit operation. For instance, the spin swap operation  has been demonstrated  with the $|T^{0}\rangle-|S\rangle$ states \cite{Spin-Qubits-Petta}, and $|T^{0}\rangle-|S\rangle$ has been utilized to implement a spin beam-splitter \cite{Spin-Beam-Splitter-Petta}. Our finding implies that the back-action should have less impact on spin qubits based on the former mechanism.

For the odd number of electrons, at zero magnetic field the two spin states are energy degenerate and are not distinguishable. Now with a parallel magnetic field we can study them separately. Fig. \ref{Figure3} (a) shows $\Gamma^{total}$ for $1e$ as a function of magnetic field. The splitting of the spin down state is overall weak, and only visible at $B\geq6.5T$.  Nonetheless, in Fig. \ref{Figure3} (b) we extracted the energy separation between $\mid\uparrow\rangle$ and $\mid\downarrow\rangle$ states at large field, which agrees well with the predicted Zeeman splitting. This verified the spin nature of these two states. Fig. \ref{Figure3} (c) and (d) show the 3D pictures for the numerical derivative of $\Gamma^{out}$ and $\Gamma^{in}$. Although noisy, the linear splitting of the two spin states with magnetic field is not missed, even at small field. The weak signal for the spin down states is possibly due to the large thermal energy ($T=240mK$). Unlike the separation of $|T^{+}\rangle$ from $|S\rangle$ by a large exchange energy, the two spin states for a single electron are too close to each other, unless at extremely large magnetic field. Thus it is not easy to distinguish the two spin states for odd electron numbers.

In conclusion, the detailed spin states spectroscopy for one and two electrons in a single QD was studied under strong back-action condition. The real-time charge counting statistics could detect the excitation from a spin up to down state for a single electron, and the excitation from spin singlet to different triplet states. We revealed that the spin exchange interaction linearly decreases with magnetic field, and found that the QPC back-action mainly drives $|S\rangle$ to the $|T^{+}\rangle$ state.

This work was supported by the NFRP 2011CBA00200 and 2011CB921200, NNSF 10934006, 11074243, 10874163, 10804104, 60921091.

\end{document}